\documentclass[a4paper]{article}
\usepackage{graphicx}
\usepackage{amsmath}
\usepackage{amssymb}
\usepackage[utf8]{inputenc}
\usepackage{hyperref}

\title{\bf Species Trees Forcing Parsimony to Fail}

\author{ 
{\bf Vikenty Mikheev }\\ Math. Dept., Kansas State University
\\ Manhattan KS 66506,USA \\ vikentym@ksu.edu 
\and
{\bf Serge E. Miheev } \\ Appl. math Dept.\\  St.Petersburg State University\\
 St.Petersburg 198504, Russia \\  him2@mail.ru
 }
 
\begin{document}
%

%

%
\maketitle              
\begin{abstract}
 To the known fact that Parsimony method sometimes fails on the problem of inferring species trees from gene trees, here we proved that no mater of what topology the true 9-taxon and greater species tree is the only thing one needs to break down Parsimony is to have in this species tree three consecutive inner edges not going through the root but perhaps ending on it with lengths $T1,T2,T3$ of some proportions. Obviously, the probability to meet these lengths is growing in general with the size of species tree.
Therefore, Parsimony may be applied only when the described lengths of edges cannot be met in the tree.
\end{abstract}

    {\bf Key Words:} {\it MRP; parsimony; coalescence theory; evolutionary trees;
     caterpillar tree; caterpillar subtree; caterpillar measure;
     caterpillar score; revolution number.}

\section{Introduction}
This article is an extended version of \cite{Miheevs} (\url{http://ceur-ws.org/Vol-2254/10000206.pdf}).

Constructing evolutionary species trees is one of the most interesting problems in biology.
It means finding the relations and mutual ancestors of existing and extinct species and also the time of formation of new species. 
Paleontology itself gives very poor information of species trees structure and time lengths of their edges. 
More precisely species trees can be built by analysis of genomes of species. 
One considers the set of species $\{a_i\}_1^N$ and their gene groups $\{G^j\}_1^K$,
 where $G^j=\{A^j_i\}_{i=1}^N$ is a set of some functionally relative to each other genes. 
For example, one group can be responsible for hemoglobin production, another one can define the eye color and so on.
 Let the gene $A^j_i$ be discovered in the species $a_i$. Then in each $j$-th functional group one can establish the relations 
between the genes in the form of unrooted tree, where   the leaves are the set of elements of $j$-th group.
 The structure of such trees for different groups can coincide (i.e. be topologically identical) or do not coincide. 
The number of these coincidences defines the frequency of the corresponding gene tree. These  frequencies are the base of parsimony method to construct evolutionary 
trees which sometimes gives wrong results.

If it is known that a method being applied to some type of problems may fail, why would anyone still use it in this area? Well, 
in phylogenetics most methods give probabilistic answers. Therefore, getting sometimes wrong answers doesn't necessarily imply 
that method is bad. To make a final conclusion about the quality of the method one could estimate how often the wrong results appear.
 Then one would compare the obtained frequency with the frequencies of other methods' failures. Having this information on the table,
 a researcher can decide if the method is acceptable.
 However, we did better than this. We have found the set of all combinations of parameters of $5$-taxon species tree
 when Parsimony is guaranteed to fail. Why tree with 5 taxa?
 It is known that Parsimony always gives right answers on $k$-taxon species tree for $k=3,4$.
 So, considering $k=5$ is quite logical from computational point of view.
 Also the smaller $k$ when things go bad, the louder the warning. 

The phylogenetics society has intuitive tendency to use Parsimony less and less. 
Nevertheless, many biologists still do it because of simplicity of the method. 
They should not be judged for that since simplicity is a strong argument.
 The results of this paper will show them the danger of Parsimony. However, forewarned is forearmed.
 If a researcher is sure that their resulting species tree doesn't have the combination of parameters we showed to be bad,
 they can safely use fast and simple Parsimony on 5-taxon trees.

For a specific rooted species tree with the known time lengths of the edges, 
using Coalescence method one can obtain the probabilities of gene trees. 
That allows to find analytically the rooted species tree and region of lengths of its edges, when parsimony fails.

\section{Preliminaries}
We consider an evolution tree $T$ (here and after we mean binary tree) of 5 species $a$, $b$, $c$, $d$
 and $e$ with some parameters $T_1$, $T_2$, $T_3$ -- time in coalescence units between the branching points (see an example in fig. \ref{T5Coal}).


Based on the Coalescent model \cite{Rosenberg}, \cite{SempleSteel}, \cite{Wakeley}, \cite{Baum}  the program COAL \newline \cite{Degnan,WangDegnan} 
yields the probabilities of all 15 possible unrooted gene trees for 5 genes $A$, $B$, $C$, $D$, $E$, 
such that $A$ is discovered in species $a$, $B$ is discovered in the species $b$ and so on.

All these species have related genes $A$, $B$, $C$, $D$, $E$, respectively. The genes originated from each other or from mutual ancestor.
The branching in gene and species trees may not coincide. That creates a problem of inferring species trees from gene trees.

Also, for 5 species there are 15 different unrooted species trees or 105 rooted ones. In each species tree, one can fit any of 15 different gene trees. However, the amount of mutations needed for this fitting will differ generally from one gene tree to another.
So, for each gene tree from these 15 and each species tree from the same 15 species trees one can correspond some non-negative integer number of mutations (parsimony score). These can be written in a $15\times 15 $ matrix $M$, where rows correspond to species trees and columns correspond to gene trees.

If one knows the frequencies of different gene trees $P=(p_1,...,p_{15})^T$, then the mathematical expectation of the number of mutations for each of 15 possible species trees can be calculated by multiplying the matrix $M$ by the vector-column $P$.

One can assume that the most probable species tree for the given sample of gene trees (or 15 gene trees with assigned probabilities) corresponds to the minimal expectation of mutations. This is the idea of Parsimony method \cite{Allman}, its application to the problem of inferring species trees from gene trees is called Matrix Representation with Parsimony (MRP) \cite{WangDegnan}.

\def\bax{\hskip 0mm {{\vbox
			{\centering 
	\hspace*{0.6cm}
	\includegraphics[scale=0.6]{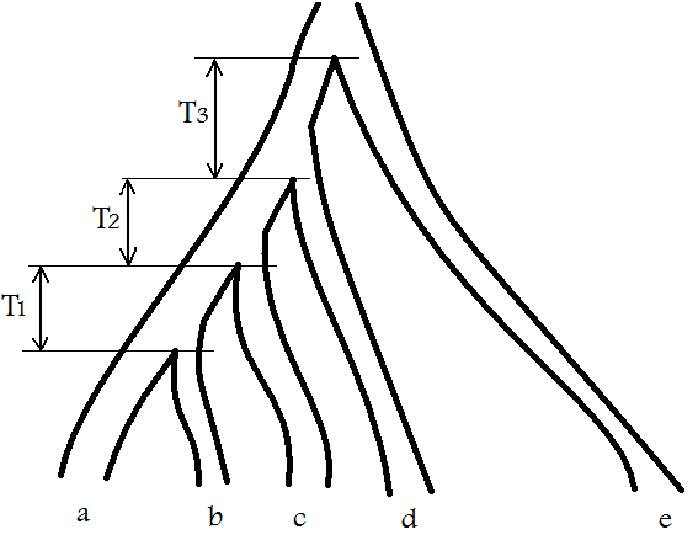}
	 \caption{ \small 5-taxon species tree of caterpillar topology.}
	\label{T5Coal}
                }}}
	
}

\
\def\baxx{\hskip 10mm {{\vbox
			{\boxmaxdepth=0mm \hsize=80.0mm \tolerance=1600
Here is the main question: Does the species tree with the minimal expectation from $M*P$, where $P$ is the vector of probabilities obtained, 
for example, from the Coalescent model \cite{Degnan} with given $T_1$, $T_2$, $T_3$ for the tree $T$, present the unrooted version of the original rooted tree $T$?

Note that we compare rooted species tree with unrooted  species tree. It is because the Coalescent method works with rooted trees 
while parsimony gives only unrooted ones.

		It appeared in our work that on the sample of gene trees obtained from a caterpillar species tree with some 
		parameters $T_1$, $T_2$, $T_3$ by coalescence, the parsimony method gives an incorrect species tree.
                }}}
}

 \begin{figure}[ht]
\hskip-45mm \bax 
\vskip-65mm\hskip75mm \baxx
\end{figure}

\section{	Numerical Experiments. Performance of Unrooted MRP for 5-Taxon Species Tree Inference.}

Any 5 genes can be joined in one unrooted tree in 15   ways as given in the following list: 

\begin{table}[ht] 
	\caption{15 topological 5-taxon trees.} 
\begin{center}
	\begin{tabular}{c}
		$\tau_1$: $((B,C),A,(D,E))$, \ \
		$\tau_2$: $((C,D),A,(B,E))$, \ \
		$\tau_3$: $((C,E),A,(B,D))$,\\
		
		$\tau_4$: $((A,E),B,(C,D))$, \ \
		$\tau_5$: $((A,D),B,(C,E))$, \ \
		$\tau_6$: $((A,C),B,(D,E))$,\\
		
		$\tau_7$: $((A,B),C,(D,E))$, \ \
		$\tau_8$: $((A,D),C,(B,E))$, \ \
		$\tau_9$: $((A,E),C,(B,D))$, \\
		
		$\tau_{10}$: $((A,B),D,(C,E))$, \ \
		$\tau_{11}$: $((A,C),D,(B,E))$, \ \
		$\tau_{12}$: $((A,E),D,(B,C))$,\\

		$\tau_{13}$: $((A,B),E,(C,D))$, \ \
		$\tau_{14}$: $((A,C),E,(B,D))$, \ \
		$\tau_{15}$: $((A,D),E,(B,C))$.\\
	\end{tabular}
	\label{Tb_ListOf15Trees}
\end{center}
\end{table}

Each unrooted tree may be transformed into a rooted tree by introducing a root to an edge. 
As a result, we have 7 rooted versions for each unrooted tree.

{\bf Step 1.} Compute parsimony scores  by Fitch-Hartigan \cite{Hartigan} 

Thus, if $\vec{N}=(N_1,N_2,...,N_{15})^{\rm\tiny T}$ is the vector of counts of 15 topological trees in the input,
$M$ is the matrix of entries in Table \ref{tablepars15_15} and vector-column \
$\vec{S}=(pars(\sigma_1),\ pars(\sigma_2),..., pars(\sigma_{15}))^{\rm\tiny T}$ then $\vec{S}=M\vec{N}$.
Here $(pars(\sigma_i)$ is parsimony score of species tree on the collection of gene trees 
$\tau_1,...,\tau_{15}$.

{\bf Step 2.} Pick the smallest entry or entries in $\vec{S}$ to determine the most parsimonious tree(s).

To study the 5-taxon case further we need to use Coalescent Theory.
The coalescent model, introduced by Kingman in \cite{Kingman}, describes the coalescence of lineages as we move backwards in time within a single species 
(Note that in biology the understanding of the word `species'  may vary. Here we use this word in the same meaning as `population'). 
By ``gluing" together such species or populations to form a tree, one gets the Multi-species Coalescent Model, which describes 
the production of gene trees within species trees.

\newpage
\begin{table}[ht]
\caption{The parsimony scores $pars_{\tau_j}(\sigma_i)\equiv m_{ij}$ for all 15 possible output trees
 $\sigma$ with respect to the matrix representation of all 15 possible input trees $\tau$.}		
	\begin{center}
	\begin{tabular}{||c || c | c |c |c |c |c |c |c |c |c |c |c |c |c |c ||}
		\hline\hline 
& $\tau_1$ & $\tau_2$ & $\tau_3$ & $\tau_4$ & $\tau_5$ & $\tau_6$ & $\tau_7$ & $\tau_8$ & $\tau_9$ & $\tau_{10}$ & $\tau_{11}$ & $\tau_{12}$ & $\tau_{13}$ & $\tau_{14}$ & $\tau_{15}$ \\ [0.5ex]
		\hline 
		$\sigma_1   $ & 2 & 4 & 4 & 4 & 4 & 3 & 3 & 4 & 4 & 4 & 4 & 3 & 4 & 4 & 3   \\ [1ex]
		$\sigma_2   $ & 4 & 2 & 4 & 3 & 4 & 4 & 4 & 3 & 4 & 4 & 3 & 4 & 3 & 4 & 4  \\ [1ex]
		$\sigma_3   $ & 4 & 4 & 2 & 4 & 3 & 4 & 4 & 4 & 3 & 3 & 4 & 4 & 4 & 3 & 4  \\ [1ex]
		$\sigma_4   $ & 4 & 3 & 4 & 2 & 4 & 4 & 4 & 4 & 3 & 4 & 4 & 3 & 3 & 4 & 4  \\ [1ex]
		$\sigma_5   $ & 4 & 4 & 3 & 4 & 2 & 4 & 4 & 3 & 4 & 3 & 4 & 4 & 4 & 4 & 3  \\ [1ex]
		$\sigma_6   $ & 3 & 4 & 4 & 4 & 4 & 2 & 3 & 4 & 4 & 4 & 3 & 4 & 4 & 3 & 4   \\[1ex]
		$\sigma_7   $ & 3 & 4 & 4 & 4 & 4 & 3 & 2 & 4 & 4 & 3 & 4 & 4 & 3 & 4 & 4  \\ [1ex]
		$\sigma_8   $ & 4 & 3 & 4 & 4 & 3 & 4 & 4 & 2 & 4 & 4 & 3 & 4 & 4 & 4 & 3  \\ [1ex]
		$\sigma_9   $ & 4 & 4 & 3 & 3 & 4 & 4 & 4 & 4 & 2 & 4 & 4 & 3 & 4 & 3 & 4  \\ [1ex]
		$\sigma_{10}$ & 4 & 4 & 3 & 4 & 3 & 4 & 3 & 4 & 4 & 2 & 4 & 4 & 3 & 4 & 4  \\ [1ex]
		$\sigma_{11}$ & 4 & 3 & 4 & 4 & 4 & 3 & 4 & 3 & 4 & 4 & 2 & 4 & 4 & 3 & 4   \\[1ex]
		$\sigma_{12}$ & 3 & 4 & 4 & 3 & 4 & 4 & 4 & 4 & 3 & 4 & 4 & 2 & 4 & 4 & 3  \\ [1ex]
		$\sigma_{13}$ & 4 & 3 & 4 & 3 & 4 & 4 & 3 & 4 & 4 & 3 & 4 & 4 & 2 & 4 & 4 \\ [1ex]
		$\sigma_{14}$ & 4 & 4 & 3 & 4 & 4 & 3 & 4 & 4 & 3 & 4 & 3 & 4 & 4 & 2 & 4  \\ [1ex]
		$\sigma_{15}$ & 3 & 4 & 4 & 4 & 3 & 4 & 4 & 3 & 4 & 4 & 4 & 3 & 4 & 4 & 2  \\[1ex] 
		\hline	\hline 
	\end{tabular}
	\end{center}
	\label{tablepars15_15} 
	
\end{table}

	Let  $g_{ij}(T)$ denote the probability that $i$ lineages (genes) since time $0$ have coalesced to exactly $j$
	lineages at time $T$ under the coalescent model. 

\medskip
General formulas for the $g_{ij}(T)$ were derived in \cite{Tavare}:
\begin{equation} \label{generalformulas}
g_{ij}(T)=\sum_{k=j}^ie^{-k(k-1)T/2}\frac{(2k-1)(-1)^{k-j}j_{(k-1)}i_{[k]}}{j! (k-j)!i_{(k)}},
\end{equation}
where \ $a_{(k)}=a(a+1)\cdots (a+k-1)$ \ for \ $k\ge 1$ \ with \ $a_{(0)}=1$ \ (the partial permutation); and\break
$a_{[k]}=a(a-1)\cdots (a-k+1)$ \ for \ $k\ge 1$ with $a_{[0]}=1$.

Some of these formulas for small indexes are

\smallskip
$	g_{11}(T)=1,
\qquad	g_{21}(T)=1-e^{-T},
\qquad  g_{22}(T)=e^{-T},
$

	$g_{31}(T)=1-(3/2)e^{-T}+(1/2)e^{-3T},
\qquad g_{32}(T)=(3/2)e^{-T}-(3/2)e^{-3T}, 
\qquad	g_{33}(T)=e^{-3T}, $

$	g_{41}(T)=1-(9/5)e^{-T}+e^{-3T}-(1/5)e^{-6T}, 
\qquad	g_{42}(T)=(9/5)e^{-T}-3e^{-3T}+(6/5)e^{-6T},$

$	g_{43}(T)=2e^{-3T}-2e^{-6T},\qquad	g_{44}(T)=e^{-6T}
$

\smallskip
Let 3 rooted species tree be \ $\Sigma_1:=((((a,b):T_1,c):T_2,d):T_3,e)$, \
$\Sigma_2:=(((a,b):T_1,(c,d):T_2):T_3,e)$ and \ $\Sigma_3:=(((a,b):T_1,c):T_2,(d,e):T_3)$. \ They are rooted versions of $\sigma_7, \sigma_{13}$ and $\sigma_7$ again, respectively (see Fig. \ref{Fcoal3r5taxa}).

\def\baxy{\hskip 0mm {{\vbox
			{\centering
	\includegraphics[width=8cm,keepaspectratio]{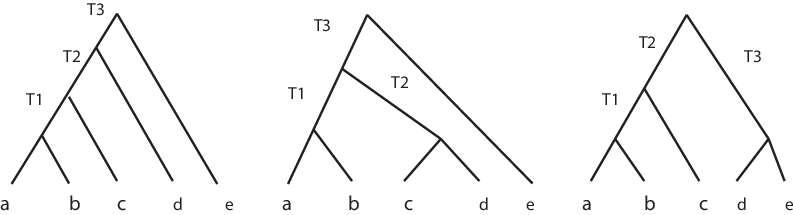}
\caption{Three rooted 5-taxon species trees $\Sigma_1, \Sigma_2$ and $\Sigma_3$.}
\label{Fcoal3r5taxa}

                }}}
	
}
\
\def\baxxy{\hskip 10mm {{\vbox
			{\boxmaxdepth=0mm \hsize=65.0mm \tolerance=1600
            Up to taxon names, these three are the only possible species trees.
They are usually referred to  as the {\it caterpillar} ($\Sigma_1$), {\it pseudo-caterpillar} ($\Sigma_2$) and {\it pseudo-balanced} ($\Sigma_3$) species trees.

                }}}
}				

\vskip1mm
\begin{figure}[ht]
\hskip-35mm  \baxy \vskip-30mm\hskip90mm \baxxy

\vskip13mm
\end{figure}

\section{An experiment with caterpillar}

Using the program COAL \cite{Degnan,WangDegnan} to get probabilities of rooted gene trees and the formulas (\ref{generalformulas}) for $g_{ij}$,
we calculate the probabilities $p_i=p(\tau_i| \Sigma_1)$ for $i=\overline{1,15}$,
which  are listed in Table \ref{Tb_Probs} ($X:=e^{-T_1},\ Y:=e^{-T_2}, \ Z:=e^{-T_3},$) after simplification in Maple 15.

Let vector-column $\mathbf p$ be $(p_1,p_2,\cdots ,p_{15})^{\tiny T}$. 

\begin{table}[h]
\caption{The probabilities \ $p_i=p(\tau_i| \Sigma_1)$ \ for \ $i=\overline{1,15}$, \ where \
 $X=e^{-T_1}$, \break $Y=e^{-T_2}, \ Z=e^{-T_3}$.}
\centering 
\begin{tabular}{||c | l |}
	\hline
		
	$   p_1$ & $X/3 - X Y/3 +X Y^3/18 + X Y^3 Z^6/90$\\  [1ex]
	$	p_2$ & $X Y^3/18 +  X Y^3 Z^6/90  $\\ [1ex]
	$	p_3$ & $X Y^3/18 +  X Y^3 Z^6/90  $\\ [1ex]
	$	p_4$ & $X Y^3/18 +    X Y^3 Z^6/90  $\\ [1ex]
	$	p_5$ & $X Y^3/18 +  X Y^3 Z^6/90 $\\ [1ex]
	$   p_6$  &$X/3 -X Y/3 +X Y^3/18 +  X Y^3 Z^6/90 $\\ [1ex]
	$   p_7$  &$1 - 2 X/3 - 2 Y/3 + X Y/3 + X Y^3/18 +  X Y^3 Z^6/90 $\\[1ex]
	$   p_8$  &$X Y^3/18 +  X Y^3 Z^6/90  $\\[1ex]
	$   p_9$  &$X Y^3/18 +  X Y^3 Z^6/90   $\\ [1ex]
	$   p_{10}$ &$Y/3 -X Y/6 -X Y^3/9 +  X Y^3 Z^6/90 $\\ [1ex]
	$   p_{11}$ &$X Y/6 -X Y^3/9 +   X Y^3 Z^6/90 $\\ [1ex]
	$   p_{12}$ &$X Y/6 -X Y^3/9 +  X Y^3 Z^6/90        $\\ [1ex]
	$   p_{13}$ &$Y/3 - X Y/6 - X Y^3/18 - 2 X Y^3 Z^6/45 $\\[1ex]
	$   p_{14}$ &$X Y/6 - X Y^3/18 -   2 X Y^3 Z^6/45$\\ [1ex]
	$   p_{15}$ &$X Y/6 -X Y^3/18 - 2 X Y^3 Z^6/45 $\\[1ex] 
	\hline 
\end{tabular}
\label{Tb_Probs}
\end{table}

 We consider the product \
  $$\mathbf s^{cat}:=M\mathbf p,$$ where each entry is the expectation of parsimony score of a possible output tree
  for MRP (see (\ref{Tb_ParsVector})).

\begin{equation} \label{Tb_ParsVector}
\mathbf s^{cat}:=M\mathbf p=\left( \begin{array}{l}
3 - X/3 + 2 Y/3 + X Y/3 - X Y^3/18 - X Y^3 Z^6/90\\
4 - Y/3 -  X Y^3/18 -  X Y^3 Z^6/90\\
4 - Y/3 - X Y^3/18 -   X Y^3 Z^6/90\\
4 - Y/3 - X Y^3/18 -  X Y^3 Z^6/90\\
4 - Y/3 - X Y^3/18 -  X Y^3 Z^6/90\\
3 - X/3 + 2 Y/3 + X Y/3 -X Y^3/18 -  X Y^3 Z^6/90\\
2 + 2 X/3 + 2 Y/3 + X Y/3 - X Y^3/18 -  X Y^3 Z^6/90\\
4 - X Y/3 -X Y^3/18 -  X Y^3 Z^6/90\\
4 - X Y/3 - X Y^3/18 - X Y^3 Z^6/90\\
3 + 2 X/3 - Y/3 + X Y/6 + X Y^3/9 - X Y^3 Z^6/90\\
4 - X/3 - X Y/6 + X Y^3/9 -  X Y^3 Z^6/90\\
4 - X/3 - X Y/6 + X Y^3/9 - X Y^3 Z^6/90\\
3 + 2 X/3 - Y/3 + X Y/6 +X Y^3/18 + 2 X Y^3 Z^6/45\\
4 - X/3 - X Y/6 + X Y^3/18 + 2X Y^3 Z^6/45\\
4 - X/3 - X Y/6 + X Y^3/18 + 2X Y^3 Z^6/45
\end{array}\right).
\end{equation}

\medskip
We discover that in $\mathbf s^{cat}$ some entries are equal.
Let's denote them as following

\medskip
$\alpha^{cat}:= s^{cat}_1= s^{cat}_6=3 - X/3 + 2 Y/3 + X Y/3 - X Y^3/18 - X Y^3 Z^6/90$,

$\beta^{cat}:= s^{cat}_2=s^{cat}_3=s^{cat}_4= s^{cat}_5=4 - Y/3 -  X Y^3/18 -  X Y^3 Z^6/90$,

$\gamma^{cat}:=s^{cat}_7:=2 + 2 X/3 + 2 Y/3 + X Y/3 - X Y^3/18 -  X Y^3 Z^6/90$, \

$\delta^{cat}:=s^{cat}_8=s^{cat}_9=4 - X Y/3 -X Y^3/18 -  X Y^3 Z^6/90,$

$\epsilon^{cat}:=s^{cat}_{10}:=3 + 2 X/3 - Y/3 + X Y/6 + X Y^3/9 - X Y^3 Z^6/90$,

$\zeta^{cat}:=s^{cat}_{11}=s^{cat}_{12}=4 - X/3 - X Y/6 + X Y^3/9 -  X Y^3 Z^6/90,$

$\eta^{cat}:=s^{cat}_{13}:=3 + 2 X/3 - Y/3 + X Y/6 +X Y^3/18 + 2 X Y^3 Z^6/45$, 

 $\theta^{cat}:=s^{cat}_{14}=s^{cat}_{15}=4 - X/3 - X Y/6 + X Y^3/18 + 2X Y^3 Z^6/45$.

\smallskip The analytical comparison of these values we form in the following

\

{\bf Proposition 1.}{\it
	For any $X,Y,Z\in (0,1)$, the following inequalities hold:
	$$\gamma^{cat}<\zeta^{cat}, \ \ \ \eta^{cat}<\theta^{cat},\ \
\gamma^{cat}<\alpha^{cat}<\beta^{cat}, \ \ \gamma^{cat}<\delta^{cat},\ \ \gamma^{cat}<\epsilon^{cat}.$$
}
 {\it Proof}.
 
 	Consider the difference \ $\alpha^{cat}-\gamma^{cat}=s^{cat}_1-s^{cat}_7=-X+1.$
 		 Since \ $X\in (0,1)$, \ $\gamma^{cat}<\alpha^{cat}$. \ Similarly,
 	$\alpha^{cat}-\beta^{cat}=s^{cat}_1-s^{cat}_2=-1+XY/3-X/3+Y=(Y-1)(X+3)/3<0
 $
  implies $\alpha^{cat}<\beta^{cat}$.
 		Let's  look at \
 		$$\gamma^{cat}-\delta^{cat}=s^{cat}_7-s^{cat}_8-= -2+(2/3)XY+(2/3)X+(2/3)Y.
     $$
      Since $X,Y$ and $X,Y$ have upper limit 1, $\gamma^{cat}-\delta^{cat}<0$.
 	Observe that \
 $$\gamma^{cat}-\epsilon^{cat}=s^{cat}_7-s^{cat}_{10}=
 -1+XY/6-XY^3/6+Y=-(Y-1)(XY^2+XY-6)/6<0.$$
  	Thus, $\gamma^{cat}<\epsilon^{cat}$.	\ Consider \
 	$	\gamma^{cat}-\zeta^{cat}=s^{cat}_7-s^{cat}_{11}=-2+XY/2-XY^3/6+X+2Y/3	=$
    $$=-2+XY(1-Y^2/3)/2+X+2Y/3<
 -2+(1/2)XY(2/3)+X+2Y/3	<-2+2/6+1+2/3 =0.
 $$
 	Thus, \     $\gamma^{cat}<\zeta^{cat}$. \
 	Finally, 
 	$$\eta^{cat}-\theta^{cat}=s^{cat}_{13}-s^{cat}_{14}=
 -1+XY/3+X-Y/3=(Y+3)(X-1)/3<0, \ {\rm i.e.} \ \eta^{cat}<\theta^{cat}.\ \
 $$
 $\square$
\medskip

\vskip5mm	However, the 3D-graphs on Figures \ref{3Dgraph_cat} show that the expressions $\eta^{cat}$ and $\gamma^{cat}$ can not be put in one order for all $X,Y,Z \in (0,1)$.

	 There is a large region were $\gamma^{cat}<\eta^{cat}$ but, nevertheless, there is also a region where $\eta^{cat}<\gamma^{cat}$. The last defines the parameters $T_1,T_2,T_3$ where MRP will fail to recover the tree topology of the true species tree producing the gene tree distribution,  even when given an arbitrary large sample of gene trees. Figure \ref{3Dgraph_cat}.left shows that provided $Y$ is not too large, regardless of $X,Z$, MRP will return the correct species tree. 
     
\vskip-3mm
\def\bax{\hskip 0mm {{\vbox
			{\boxmaxdepth=0mm \hsize=90.0mm \tolerance=1600
				\includegraphics*[width=70mm, height=70mm, keepaspectratio=false]{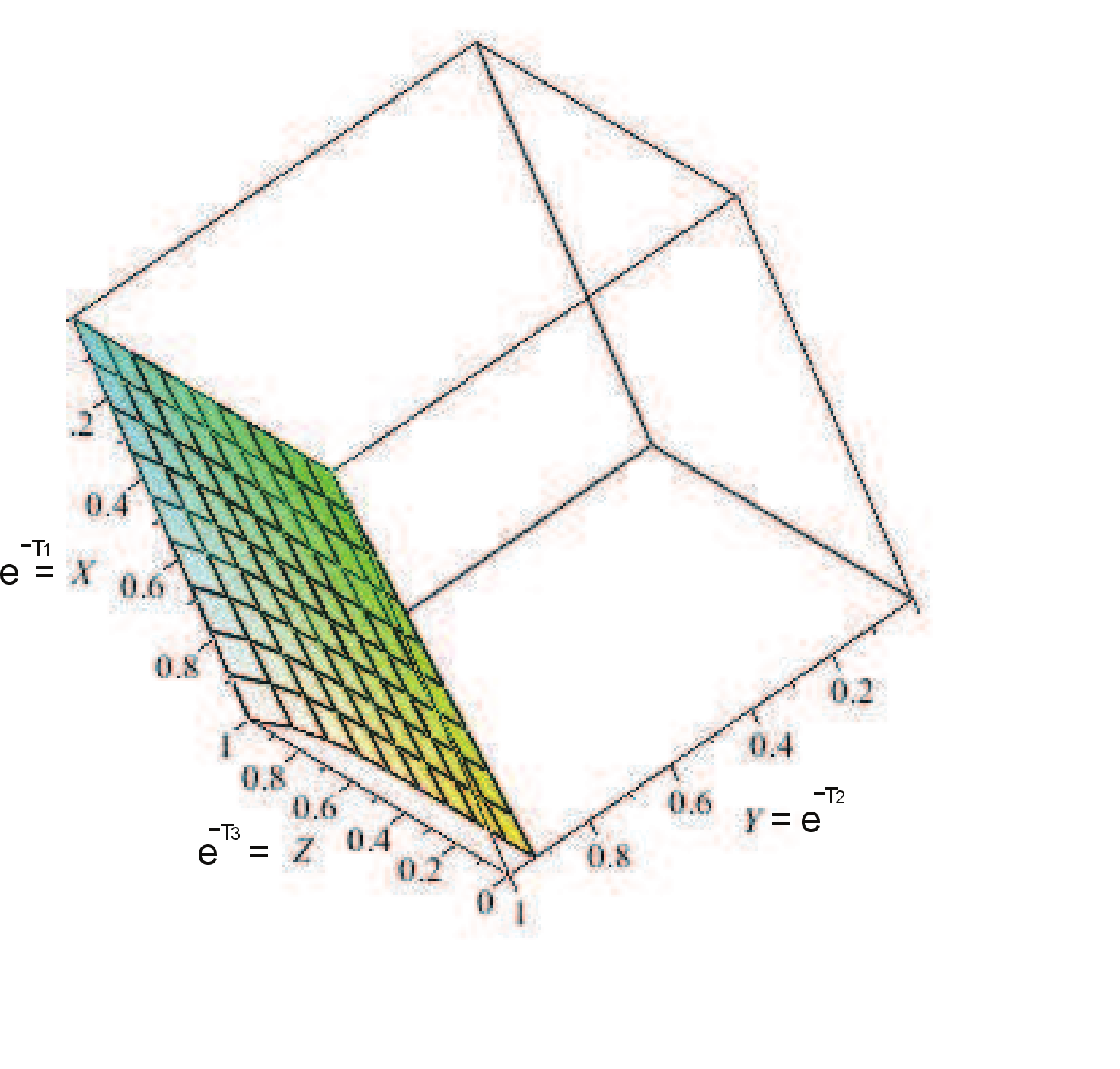} }}}/images
}

\
\def\baxx{\hskip 10mm {{\vbox
			{\boxmaxdepth=0mm \hsize=80.0mm \tolerance=1600
				\includegraphics*[width=60mm, height=60mm,
				keepaspectratio=false]{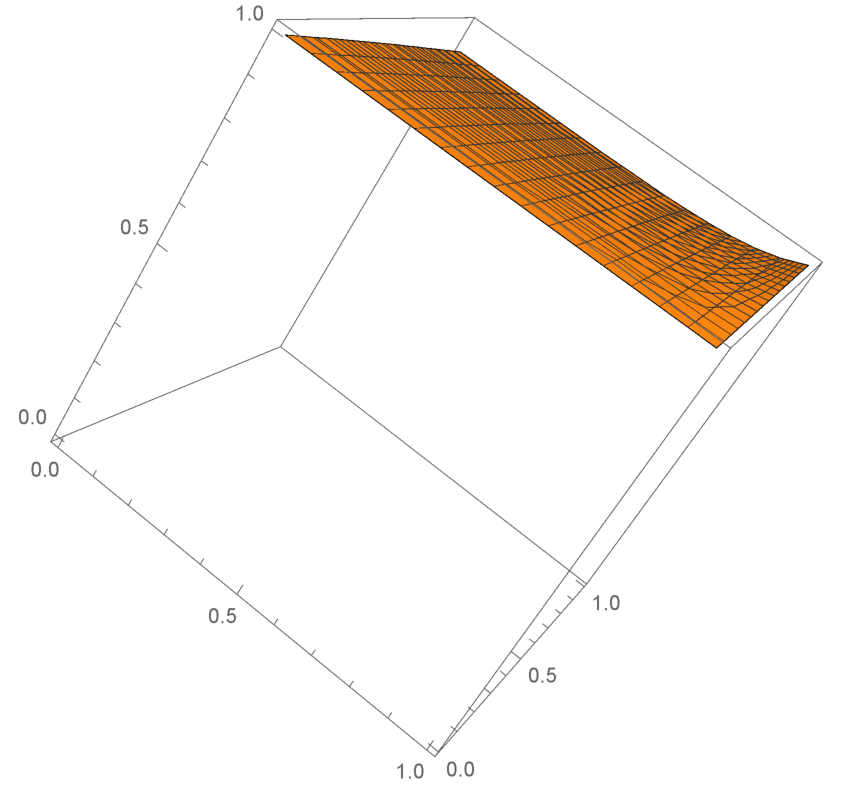}}}}
}				
 \begin{figure}[h]
\hskip31mm $ \bax$\vskip-70mm\hskip80mm$ \baxx$

\vskip3mm
\caption{Left: the surface $\eta^{cat}(X,Y,Z)=\gamma^{cat}(X,Y,Z)$. $\eta^{cat}>\gamma^{cat}$ on the large region including all those points, when $Y$ is near 0, while $\eta^{cat}<\gamma^{cat}$ on the small region. Right: Different angle on the same surface $\eta^{cat}(X,Y,Z)=\gamma^{cat}(X,Y,Z)$. }
\label{3Dgraph_cat}
\end{figure}


To determine this cutoff for $Y$, we set $\eta^{cat}(1,Y,0)=\gamma^{cat}(1,Y,0)$ 
and solve to get $Y=0.935...$ (the solutions of \ $(1/3)Y^3-(7/2)Y+3 = 0$ are \
$Y_{1,2,3}\approx 2.670...,-3.605...,  0.935...$).

		\section{An experiment with  pseudo-caterpillar}
	Let the pseudo-caterpillar species tree be $\Sigma_{2}=(((a,b):T_1,(c,d):T_2):T_3,e)$.
	We use COAL, the formulas for $g_{ij}$ and Maple 15 to calculate the probabilities $p^{pc}_i=p(\tau_i| \Sigma_2)$ for $i=\overline{1,15}$. 
	Their list is shown in Table \ref{Tb_Probs_2} after simplifications.
\begin{table}[h]
\caption{The probabilities $p^{pc}_i=p(\tau_i| \Sigma_2)$ for $i=\overline{1,15}$.}
\centering 
\begin{tabular}{|c l |}
	\hline		
&\\[1ex]
		$p^{pc}_1$&$=p^{pc}_3=p^{pc}_5=p^{pc}_6=p^{pc}_8=p^{pc}_9= p^{pc}_{11}=p^{pc}_{12}=(1/18)XY+XYZ^6/90,$\\[1ex]
		$p^{pc}_2$&	$=p^{pc}_4=p^{pc}_7=	p^{pc}_{10}=XYZ^6/90+Y/3-(5/18)XY$, \\[1ex]
		$p^{pc}_{13}$& $=1-(2/45)XYZ^6+(4/9)XY-(2/3)X-(2/3)Y,$ \\[1ex]
		$p^{pc}_{14}$& $=	p^{pc}_{15}=-(2/45)XYZ^6+XY/9.$\\[1ex]
\hline 
\end{tabular}
\label{Tb_Probs_2}
\end{table}

\noindent      In the table \ref{Tb_Probs_2}: $X=e^{-T_1}$, \ $Y=e^{-T_2}, \ Z=e^{-T_3}$


Now assuming \ ${\mathbf p}^{pc}=(p_1^{pc},p_2^{pc},\cdots ,p_{15}^{pc})^{\tiny T}$ \ we see that the product \
  $\mathbf s^{pc}:=M\mathbf p^{pc}$ (see (\ref{Tb_ParsVector2})) is the vector of expected parsimony scores of possible  output trees with a pseudo-caterpillar species tree. 
\begin{equation} 		\label{Tb_ParsVector2}
\mathbf s^{pc}:=M\mathbf p^{pc}=\left(\begin{array}{l}
4- X Y Z^6/90- Y/3- X Y/18 \\
3- X Y Z^6/90-X/3+2Y/3+5X Y/18\\
4-X Y Z^6/90- Y/3-X Y/18\\
3-X Y Z^6/90- X/3+2Y/3+5 X Y/18\\
4-X Y Z^6/90- Y/3- X Y/18\\
4-X Y Z^6/90- Y/3-X Y/18\\
3-X Y Z^6/90+2X/3-Y/3+5 X Y/18\\
4-X Y Z^6/90-X/3- X Y/18\\
4-X Y Z^6/90- X/3- X Y/18\\
3-X Y Z^6/90+2X/3-Y/3+5X Y/18\\
4-X Y Z^6/90- X/3- X Y/18\\
4-X Y Z^6/90- X/3-X Y/18\\
2+2 X Y Z^6/45+2X/3+2Y/3+2X Y/9\\
4+2X Y Z^6/45-4X Y/9\\
4+2 X Y Z^6/45-4X Y/9
\end{array}\right).
\end{equation}

We discover that in $\mathbf s^{pc}$ some entries are equal.
Let's denote them as following

\medskip 
$\alpha^{pc}:=s^{pc}_1=s^{pc}_3=s^{pc}_5=s^{pc}_6=4- X Y Z^6/90- Y/3-XY/18 $,

\medskip  $\beta^{pc}:=s^{pc}_2=s^{pc}_4=3- X Y Z^6/90-X/3+2Y/3+5X Y/18$, 

\medskip  $\gamma^{pc}:=s^{pc}_7=s^{pc}_{10}=3-X Y Z^6/90+2X/3-Y/3+5 X Y/18$, 

\medskip  $\delta^{pc}:=s^{pc}_8=s^{pc}_9=s^{pc}_{11}=s^{pc}_{12}=4-XYZ^6/90-X/3-X Y/18$,
 
\medskip    $\zeta^{pc}:=s^{pc}_{13}=2+2 X Y Z^6/45+2X/3+2Y/3+2X Y/9$, 
  
\medskip    $\eta^{pc}:=s^{pc}_{14}=s^{pc}_{15}=4+2X Y Z^6/45-4X Y/9$. 
  
\medskip  \noindent 	Let's compare these expressions.

\smallskip    
	{\bf Proposition 2.} \ {\it
		For any $X,Y,Z\in (0,1)$, the following inequalities hold:
		$$ \zeta^{pc}<\alpha^{pc}, \ \zeta^{pc}<\beta^{pc},\ \zeta^{pc}<\gamma^{pc}, \ \zeta^{pc}<\delta^{pc}, \ \zeta^{pc}<\eta^{pc}.$$
    }
{\it Proof.} 
		Consider the difference $$\alpha^{pc}-\zeta^{pc}=s^{pc}_1-s^{pc}_{13}=2-XYZ^6/18-Y-5XY/18-2X/3.
 $$
 It is easy to see that, since $X,Y,Z\in (0,1)$, \ $\alpha^{pc}-\zeta^{pc}>0$, i.e. $\alpha^{pc}>\zeta^{pc}$.
		
		Observe that \ $	\beta^{pc}-\zeta^{pc}= s^{pc}_2-s^{pc}_{13}=$
 $$1-XYZ^6/18-X+XY/18=	1-X[YZ^6/18+1-Y/18]>
 $$$$1-YZ^6/18-1+Y/18=		Y[-Z^6+1]/18>0.
		$$
 		Thus, $\beta^{pc}>\zeta^{pc}$.
 				Similarly, \ $	\gamma^{pc}-\zeta^{pc}=s^{pc}_7-s^{pc}_{13}=$
 $$				1-XYZ^6/18-Y+XY/18=1-Y[XZ^6/18+1-X/18]>
 		$$$$		>-XZ^6/18+X/18=X[-Z^6+1]/18>0.
 		$$
 		Thus, $\gamma^{pc}>\zeta^{pc}$.		Observe $$\delta^{pc}-\zeta^{pc}=s^{pc}_8-s^{pc}_{13}=2-(1/18)XYZ^6-X-(5/18)XY-(2/3)Y>0\  {\rm and}
 $$$$\eta^{pc}-\zeta^{pc}=s^{pc}_{14}-s^{pc}_{13}=2-(2/3)XY-(2/3)X-(2/3)Y>0, \ \ {\rm so}\ \
 \delta^{pc}>\zeta^{pc}$$
 and  \ $\eta^{pc}>\zeta^{pc}. $ \ \ \
 $\square$
  
  \medskip

This implies that if the true species tree is 5-taxon pseudo-caterpillar,
 MRP, for a sufficiently large data set, will give with probability 1
the unrooted species tree topology for all $T_1,T_2,T_2\in (0,1)$.

	\section{An experiment with	psuedo-balanced}

	The last tree we need to consider (since all other are just permutations of taxon names)
 is the pseudo-balanced species tree $\Sigma_3=(((a,b):T_1,c):T_2,(d,e):T_3)$.

 	The same chain of actions give us the probabilities $p^{pb}_i$ (Table \ref{Tb_Probs_3_1unique}).
\begin{table}[h]
 	\begin{center} 
 	\caption{The probabilities $p^{pb}_i=p(\tau_i| \sigma_{7})$ for $i=
 				\overline{1,15}$.}	
 	\begin{tabular}{|r l |}
 	\hline 
 	$p^{pb}_1$ &$=p^{pb}_6=-(1/3)ZXY+(1/15)XY^3Z+(1/3)X,$\\
 		$p^{pb}_2$&$=p^{pb}_3=p^{pb}_4=p^{pb}_5=(1/15)XY^3Z,$\\
 		$p^{pb}_7$&$=1+(1/3)ZXY-(2/3)X+(1/15)XY^3Z-(2/3)ZY,$ \\
 		$p^{pb}_8$&$=p^{pb}_9=(1/15)XY^3Z,$\\
 		$p^{pb}_{10}$&$=p^{pb}_{13}=(1/3)ZY-(1/6)ZXY-(1/10)XY^3Z,$ \\
 		$p^{pb}_{11}$&$=p^{pb}_{12}=p^{pb}_{14}=p^{pb}_{15}=(1/6)ZXY-(1/10)XY^3Z.$\\
 	\hline	\end{tabular}
 								\label{Tb_Probs_3_1unique}
 		
 	\end{center}\end{table}
		
 	Again,  multiplying the matrix $M$  by the vector $\mathbf p^{pb}:=(p^{pb}_1,p^{pb}_2,\cdots ,p^{pb}_{15})$,
  gives the vector of expected parsimony scores of possible output trees
 	 (\ref{Tb_ParsVector3}) with a pseudo-balanced species tree $\Sigma_3$ as input.

 	\begin{equation}
 		\mathbf s^{pb}:=M\mathbf p^{pb}=\left(\begin{array}{l}
 		3+(1/3)ZXY-(1/3)X-(1/15)XY^3Z+(2/3)ZY\\
 		4-(1/15)XY^3Z-(1/3)ZY\\
 		4-(1/15)XY^3Z-(1/3)ZY \\
 		4-(1/15)XY^3Z-(1/3)ZY \\
 		4-(1/15)XY^3Z-(1/3)ZY \\
 		3+(1/3)ZXY-(1/3)X-(1/15)XY^3Z+(2/3)ZY\\
 		2+(1/3)ZXY+(2/3)X-(1/15)XY^3Z+(2/3)ZY \\
 		4-(1/3)ZXY-(1/15)XY^3Z \\
 		4-(1/3)ZXY-(1/15)XY^3Z \\
 		3+(1/6)ZXY+(2/3)X+(1/10)XY^3Z-(1/3)ZY\\
 		4-(1/6)ZXY-(1/3)X+(1/10)XY^3Z \\
 		4-(1/6)ZXY-(1/3)X+(1/10)XY^3Z \\
 		3+(1/6)ZXY+(2/3)X+(1/10)XY^3Z-(1/3)ZY \\
 		4-(1/6)ZXY-(1/3)X+(1/10)XY^3Z \\
 		4-(1/6)ZXY-(1/3)X+(1/10)XY^3Z
 		\end{array}\right). 		\label{Tb_ParsVector3}
 	\end{equation}

 	Let's denote $\alpha^{pb}:=s^{pb}_1=s^{pb}_6$, \ \ $\beta^{pb}:=s^{pb}_2=s^{pb}_3=s^{pb}_4=s^{pb}_5$, \ \
  $\gamma^{pb}:=s^{pb}_7$, \ \ $\delta^{pb}:=s^{pb}_8=s^{pb}_9$,\ \ $\zeta^{pb}:=s^{pb}_{10}=s^{pb}_{13}$ \ \ and \ \
 $\eta^{pb}:=s^{pb}_{11}=s^{pb}_{12}=s^{pb}_{14}=s^{pb}_{15}$.
	
\medskip	
 	{\bf Proposition 3.} {\it
 		For any $X,Y,Z\in (0,1)$, the following inequalities hold:
 $$ \alpha^{pb}>\gamma^{pb}, \ \beta^{pb}>\gamma^{pb}, \ \delta^{pb}>\gamma^{pb}, \ \zeta^{pb}>\gamma^{pb}, \ \eta^{pb}>\gamma^{pb}.
 $$
     }
 	{\it Proof.}
 		Since $X\in (0,1)$, we immediately have that the difference $\alpha^{pb}-\gamma^{pb}$ equals to $s^{pb}_1-s^{pb}_7=1-X>0$. \ So, \ $\alpha^{pb}>\gamma^{pb}$. Also $$\beta^{pb}-\gamma^{pb}=s^{pb}_2-s^{pb}_7=2-ZY-(1/3)ZXY-(2/3)X>0
 $$ and
 $$\delta^{pb}-\gamma^{pb}=s^{pb}_8-s^{pb}_7=2-(2/3)ZXY-(2/3)X-(2/3)ZY>0.$$
 		Thus, \ $\beta^{pb}>\gamma^{pb}$ \ and \ $\delta^{pb}>\gamma^{pb}$.
		
 		Consider \ $\zeta^{pb}-\gamma^{pb}=s^{pb}_{10}-s^{pb}_7=
 		1-ZY(1+(1/6)X(1-Y^2))$
  $$>1-Y(1+X(1-Y^2)/6)>		1-Y(1+(1-Y^2)/6)=1-7Y/6+Y^3/6=:h
 		$$
		
 		Since \ $h'=-7/6+(1/2)Y^2<0$ \ and \ $h(0)<0$, \ $h>0$ for all $Y\in (0,1)$. \ Thus, \ $\zeta^{pb}>\gamma^{pb}$.
		
 		Finally, observe that \ $\eta^{pb}-\gamma^{pb}=s^{pb}_{11}-s^{pb}_7=	$
 $$			(2-x)-ZYX/2-XY^2/6+2/3)> 1-ZY(2/3+X(1-Y^2/3)/2)
 		$$$$
 		>1-Y(2/3+X(1/2-Y^2/6))>1-Y(2/3+1/2-Y^2/6)=
 		1-7Y/6+Y^3/6=h>0.
 		$$
 		So, $\eta^{pb}>\gamma^{pb}$. \ \ \ $\square$		
 		
\medskip 
	
	So, if the true species tree is 5-taxon pseudo-balanced $\Sigma_3$, MRP,
  for a sufficiently large data set, will give with probability 1 the correct
unrooted species tree topology for all $T_1,T_2,T_2\in (0,1)$.

		\section{Generalization of results.}
	
	\subsection{Caterpillar Subtree.}
	
	{\bf Definition 2.}
		There is a rooted tree $T$ with number of taxa equal to \
 $|T|=:n$. Let $T_{cat}(T)$ be a caterpillar subtree of this tree. The number\
  $\displaystyle Cat(T):= \max_{T_{cat}(T)\subset T}|T_{cat}(T)|$ for a particular tree $T$ is called {\it caterpillar score for the tree $T$}.
		The number $\displaystyle cat(n):=\min_{|T|=n}\ Cat(T)$ is called {\it caterpillar measure}.
	
	It is clear that $cat(n)$ is an increasing function with respect to $n$.
	There are may be a few consecutive numbers $n$ such that $cat(n)=k$ for some given natural $k$.
	
	{\bf Definition 3.}
		Let's call number $\displaystyle r_k:=\min_{cat(n)=k} n$ the {\it revolution number}.
	
	Observe that for the caterpillar lengths $1,2,3$ their revolution numbers are $r_1=1,r_2= 2,r_3= 3$, because these trees are caterpillar themselves.
		Note, that the third and the second revolution numbers are connected by
	\begin{equation}
	r_3=2r_2-1.
	\label{eq1}
	\end{equation}

{\bf Theorem 1.}{\it	
		The revolution numbers $r_k$ may be calculated recursively \ $r_{k}=2r_{k-1}-1$, $k=4,5...$.
    }
    
	{\it Proof.}
			Let $T$ be a $k$-taxon tree. Since $T$ is a binary tree, we can  think of $T$ as two subtrees $T',T''$ glued together only by two edges at the root (see Figure \ref{Tree1}). Observe that for any caterpillar subtree of $T'$ one of these $T',T''$ being transformed properly brings only one edge to the caterpillar.

		On the other hand, \ $r_k$ is an increasing function. So, the first bifurcation in the root will be the worst in the sense of caterpillar score for the tree $T$ when this bifurcation divides the tree into two subtrees $T'$ and $T''$ such that $|T'|=|T''|$ for even $|T|$, and $\bigl||T'|-|T''|\bigr|=1$ for odd $|T|$.
				Further, we use mathematical induction.
		
		\medskip
		{\it Base of induction. } 
				The formula (\ref{eq1}) is the case $k=2$.
				
				\medskip
				{\it Assumption of induction.}	
				Let $r_3,...,r_k$ calculated recursively be revolution numbers.
		
		\medskip
		{\it Inductive step.}		
		We need to prove that $r_{k+1}=2r_k-1$ is the revolution number. Let us take a tree $T$ with $r_{k+1}$ taxa and consider the worst bifurcation in its root.		
		As mentioned above, the worst bifurcation in the root of $T$ forms two subtrees $T'$ and $T''$ such that $|T'|=r_k-1,\ |T''|=r_k$.




\def\baxxz{\hskip 5mm {{\vbox
			{\boxmaxdepth=0mm \hsize=95.0mm \tolerance=1600
     The induction assumption yields an existence of caterpillar in $T''$ with a length no less than $k$.
     One edge in $T'$ together with the caterpillar in $T''$ creates the caterpillar tree $C$ with $|C|=k+1$.	So, the revolution number for $k+1$ is no greater than $2r_k-1$.
        If $|T|\in [r_k,2r_k-1)$, then  the worst bifurcation forms two subtrees $T'$ and $T''$ such that $|T'|,|T''|<r_k$. Since $r_k$ is a revolution number, there are $T'$ and $T''$ which have only caterpillars $C',C''$ and $C''$ with $|C'|,|C''|<r_k$. Therefore, $r_{k+1}$ is the revolution number.
   
                }}}
}

\def\baxz{\hskip 0mm {\vbox
			{\boxmaxdepth=0mm \hsize=90.0mm \tolerance=1600
\centering
\includegraphics[height=40mm]{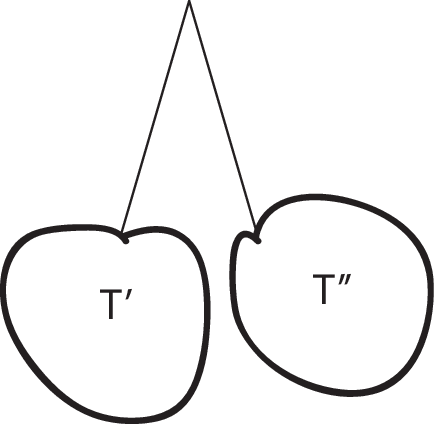}
\caption{}
\label{Tree1}
}}}

\newpage
 \begin{figure}[h]
\hskip-25mm \baxz \vskip-45mm\hskip59mm$ \baxxz$

\vskip3mm
\end{figure}

\medskip

\noindent	Theorem 1 
allows to continue the sequence of the revolution numbers  for the caterpillar measures $3,4,5,6,7,...$ as $r_k=3, 5, 9, 17,33,...$, respectively. For trees with number of taxa in $[r_k,2r_k-1]$, the caterpillar measure is $k$.

	\subsection{MRP on trees with the number of taxa greater than 5}
	
{\bf Theorem 2.}{\it
		\label{propSubTree}
		If a true species rooted tree $\hat G$ contains 5-taxon caterpillar subtree, then MRP may fail to obtain the unrooted version of $\hat G$ from the set of gene trees generated by Coalescent model from $\hat G$.
    }
    
	{\it Proof.}
				For the number of species greater than 5 and the same number of genes one can make the following construction.

		Take a caterpillar tree $\Gamma$ of 5 species $a$, $b$, $c$, $d$, $e$ with $T_1$, $T_2$, $T_3$ and root $\rho$, 
		such that parsimony fails (fig. \ref{3Dgraph_cat}). Take an arbitrary tree $G$, where every edge is very close to 0 ($T_i\approx 0$, \ $i>4$).
		Connect $\Gamma$ to $G$ through its root $\rho$ and the edge $\epsilon$ with the length $T_4$.
		Make $T_4$ big enough so the genes  $A$, $B$, $C$, $D$, $E$ coalesce in $\epsilon$ if they didn't in $\Gamma$. 
		No matter what is on the upper end of $\epsilon$, root of entire tree or inner node created by $\epsilon$ on some edge of $G$ .

		The numeration of $n$ possible gene trees we do in the following way: First 15 trees will have the same subtree $G$ and different topology or permutation of $A$, $B$, $C$, $D$, $E$. Other $n-16$ trees can be numerated in any order, and we set their probabilities $p_{16},..., p_{n}$ equal to zero, since $T_i$, \ $i\in \{5,..., n\}$, can be taken infinitively small.
		Therefore, the set of gene trees is numerated and the probabilities of them	are presented by vector-column $p=(p_1,...,p_{15},p_{16},...,p_{n})^T$, where all the entries below 15-th equal zero and the first 15 are the same that obtained ones for 5-taxon experiment.
		The matrix $M_{n\times n}$ has dimension $n\times n$, but only submatrix  $M_{n\times 15}$ does participate in calculation of expectations of gene mutations due to	
		$p_{16},...,p_{n}=0$.
		
		Moreover, the parsimony incorrect choice may be shown on submatrix $M_{15\times 15}$ in upper corner.
		Observe that the elements of $M_{15\times 15}$ are the sums of the elements of $M$ obtained earlier 
		in performance of MRP for 5-taxon trees (see table \ref{tablepars15_15})
		and some constant number generated by the constant subtree $G$ with coalesced gene $A+B+C+D+E$.
		This means that each of $s^{cat}_1$, $s^{cat}_2$, ... , $s^{cat}_{15}$ from Section 
		\ref{Tb_ParsVector} must be increased by the constant value \ \ $V \sum_1^{15} p_{i}$, \ to be a new mathematical expectation for the new big tree. So, the minimum among the first 15 rows must be achieved in the same index.
		Therefore, being wrong for 5-taxon caterpillar species tree the parsimony becomes wrong for the constructed tree $\hat G$ as well.
		\ \ \ \ $\square$

	\medskip
	{\bf Corollary 1.} {\it
		MRP on a set of gene trees with 5 taxa or more  may yield wrong result. If one applies MRP on set of gene trees with 9 taxa or more, MRP may fail even more probably, since 9-taxon species tree always has a caterpillar subtree, which may have unfortunate lengths of inner edges from the small region in Fig. \ref{3Dgraph_cat}.
    }
	
We have established what these unfortunate lengths are. But how to find these caterpillar subtrees?  One may use the following
    
{\bf Theorem 3.} {\it
If a tree has three consecutive inner edges not contaning the root betwen them but perhaps one of these edges ending on it then the tree has a 5-taxon caterpillar subtree, which contains these three inner edges.
}   

\begin{figure}[h]
\centering
\includegraphics[height=23mm]{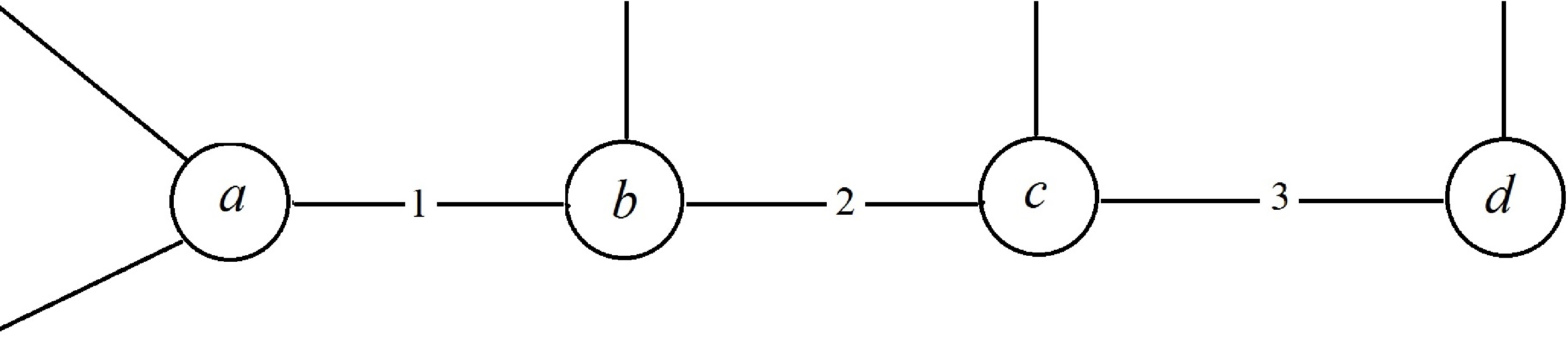}

\caption{Three consecutive inner edges in some tree.}
\label{fig:ThreeInnerEdges}
\end{figure}

{\it Proof.}
At Fig. \ref{fig:ThreeInnerEdges} we can see three consecutive edges denoted 1,2 and 3 between nodes $a$,$b$, $c$ and $d$, respectively and the third edge is closest to the root. Since these edges are inner, all the nodes must be bifurcation points and so
each of $b$, $c$, $d$ has one more edge running  towards a taxon or a clade opposite to the root and $a$ has two more such edges. 
Let $d$ be the root then simply contracting each of the mentioned clades to one of its taxa we get a 5-taxon caterpillar subtree.
If $d$ is not the root, throw away one of the edges at $d$ to make $d$ the root of the 5-taxon caterpillar subtree.

{\bf Corollary 2.}{\it
It is enough to know that in a species tree with amount of taxa 5 or more there are three consecutive inner edges not going through the root but perhaps ending on it with lengths $T_1, \ T_2$ and $T_3$ from the small region of cube in Figure \ref{3Dgraph_cat} to conclude that Parsimony is guaranteed to fail on this tree.
 }

 \section{Conclusions and Future work}

	The fact that Parsimony may fail is not new. However, here we proved that no mater of what topology the true 9-taxon and greater species tree is the only condition to fail Parsimony is to have in this tree three consecutive inner edges not going through the root but perhaps ending on it with lengths $T1,T2,T3$ (which are times in coalescence units between the branching points) of some proportions. Obviously, the probability to meet these lengths is growing in general with the size of species tree.
	    So, if one wants to safely use MRP on a set of $n$-taxon gene trees, it is need to know somehow that the resulting $n$-taxon species tree cannot have any of ``bad" topologies and edge lengths from ``bad" regions. This paper makes it possible for $n\leq 5$. Also, one may apply MRP on a set of 5-taxon gene trees and if the result is the caterpillar tree or topology $\Sigma_3$ from Fig. \ref{Fcoal3r5taxa}, it is true.


One may consider  6-taxon species trees the way we did in this paper and check the existence  of 6-taxon topology which forces Parsimony to fail when lengths of inner edges have some proportions. Then prove, perhaps following our ideas, that every tree with some number of taxa greater than 6 has this 6-taxon topology subgraph. 
Then do the same for 7 taxa and so on. 

If one studies all ``bad" parameters' regions of all ``bad" topologies for all trees with the amount of taxa less or equal some $n$, it becomes theoretically possible to check either MRP can be applied for a set of gene trees of $n$ taxa (if, of course, the researcher knows enough information about possible results). 
However, taking into account the factorial growth of the amount of binary trees with respect to the amount of taxa, the problem to find ``safe zone" for MRP becomes extremely hard. 
Unfortunately, we don't see any way around besides doing that scheme for each $k<n$. So, someone has to be very motivated to use MRP to go through with the research. Nowadays, there are good coalescence based methods, for example, \cite{Wu}, \cite{Rannala} and \cite{Emms}

It could be interesting to study stability questions of the coalescent model with uncertainty in data  applying the thoughts from \cite{Zubov}.

	\bibliographystyle{alpha}

\begin{thebibliography}{Wak08}
\bibitem[Miheevs]{Miheevs} Vikenty Mikheev and Serge E.~Miheev.
	\newblock Species trees forcing the parsimony to fail modelling evolution process.	
	\newblock {\em Proceedings of the 2018 Multidisciplinary Symposium on Computer Science and ICT (REMS 2018)}.
	\newblock Stavropol, Russia, 2018. \url{http://ceur-ws.org/Vol-2254/}

		\bibitem[AllmanRhodes]{Allman}
		E.~S. Allman and J.~A. Rhodes.
		\newblock  Lecture Notes: The Mathematics of Phylogenetics.
		\newblock {\em University of Alaska Fairbanks}, 2009.
		
		\bibitem[Baum]{Baum}
		B.~R. Baum.
		\newblock Combining trees as a way of combining data sets for phylogenetic
		inference, and the desirability of combining gene trees.
		\newblock {\em Taxon}, 41:3 -- 10, 1992.
		
		\bibitem[DegnanSalter]{Degnan}
		J.~H. Degnan and L.~A. Salter.
		\newblock Gene tree distributions under the coalescent process.
		\newblock {\em Evolution}, 59(1):24 -- 37, 2005.
		
		\bibitem[Hartigan]{Hartigan}
		J.~A. Hartigan.
		\newblock Minimum mutation fits to a given tree.
		\newblock {\em Biometrics}, 29:53 -- 65, 1973.
		
		\bibitem[Kingmn]{Kingman}
		J.~F.~C. Kingman.
		\newblock The coalescent.
		\newblock {\em Stoch. Process. Appl.}, 13:235 -- 248, 1982.
		
		\bibitem[Rosenberg]{Rosenberg}
		N.~A. Rosenberg.
		\newblock The probability of topological concordance of gene trees and species
		trees.
		\newblock {\em Theor. Pop. Biol.}, 61:225 -- 247, 2002.
		
		\bibitem[SemSte]{SempleSteel}
		Charles Semple and Mike Steel.
	\newblock {\em Phylogenetics}, volume~24 of {\em Oxford Lecture Series in
			Mathematics and its Applications}.
		\newblock Oxford Un. Press, Oxford, 2003.
		
		\bibitem[Tavar{\'e}]{Tavare}
		S. Tavar{\'e}.
		\newblock Line-of-descent and genealogical processes, and their applications in
		population genetics models.
		\newblock {\em Theoret. Population Biol.}, 26(2):119--164, 1984.
		
		\bibitem[Wakeley]{Wakeley}
		J.~Wakeley.
		\newblock {\em Coalescent Theory}.
		\newblock Roberts \& Company, Greenwood Village, CO, 2008.
		
		\bibitem[WangDegnan]{WangDegnan}
		Yuancheng Wang and James~H. Degnan.
		\newblock Performance of matrix representation with parsimony for inferring
		species from gene trees.
		\newblock {\em Stat. Appl. Genet. Mol. Biol.}, 10:Art. 21, 41, 2011.
		
        \bibitem[Zubov]{Zubov}
        I.~V. Zubov and A.~V. Zubov.
        \newblock The stability of motion of dynamic systems.
        \newblock {\em Doklady Mathematics}, 79(1):112 -- 113, 2009.
        
        \bibitem[YufengWu]{Wu}
        Yufeng Wu.
        \newblock A coalescent-based method for population tree inference with haplotypes
        \newblock {\em Bioinformatics}, 31(5):691 -- 698, 2015.
        
        \bibitem[RanYan]{Rannala}
        Bruce Rannala and Ziheng Yang.
        \newblock Efficient Bayesian Species Tree Inference under the Multispecies Coalescent
         \newblock {\em Systematic Biology}, 66(5): 823–842, 2017.
         
          \bibitem[EmmsKelly]{Emms}
           David Emms and Steven Kelly.
          \newblock STAG: Species Tree Inference from All Genes
	 \newblock {\em Biorxiv }, 2018.
	 
     \url{https://doi.org/10.1101/267914 }
          
     
    \end{thebibliography}
	



\end{document}